# Shooting method for solving two-point boundary value problems in ODEs numerically

**Jitender Singh[1]**

[1]Department of Mathematics, Guru Nanak Dev University Amritsar-143005, Punjab, INDIA
sonumaths@gmail.com; jitender.math@gndu.ac.in

**Abstract.** Boundary value problems in ODEs arise in modelling many physical situations from microscale to mega scale. Such two-point boundary value problems (BVPs) are complex and often possess no analytical closed form solutions. So, one has to rely on approximating the actual solution numerically to a desired accuracy. To approximate the solution numerically, several numerical methods are available in the literature. In this chapter, we explore on finding numerical solutions of two-point BVPs arising in higher order ODEs using the shooting technique. To solve linear BVPs, the shooting technique is derived as an application of linear algebra. We then describe the nonlinear shooting technique using Newton-Kantorovich theorem in dimension $n > 1$. In the one-dimensional case, Newton-Raphson iterates have rapid convergence. This is not the case in higher dimensions. Nevertheless, we discuss a class of BVPs for which the rate of convergence of the underlying Newton iterates is rapid. Some explicit examples are discussed to demonstrate the implementation of the present numerical scheme.

## 1. Introduction.

For a positive integer $n > 1$ and real numbers $a$ and $b$ with $a < b$, let $\boldsymbol{X}: [a, b] \to \boldsymbol{R^n}$ be the $n \times 1$ column matrix function defined as follows:

$$\boldsymbol{X}(z) = \big(X_1(z), X_2(z), \ldots, X_n(z)\big)',$$

where the prime denotes matrix transpose and $X_i: [a, b] \to \boldsymbol{R}$, for each $1 \leq i \leq n$ is sufficiently smooth on the open interval $(a, b)$. We consider the following first order ordinary vector, differential equation

$$\frac{d}{dz}\boldsymbol{X}(z) = f\big(z, \boldsymbol{X}(z)\big), z \in [a, b]. \tag{1}$$

Throughout, the map $f: [a, b] \times \boldsymbol{R}^n \to \boldsymbol{R}^n$ is assumed to be sufficiently smooth and satisfies Lipschitz condition on the closed rectangle $R = [a, b] \times \overline{U} \subset [a, b] \times \boldsymbol{R}^n$ with a Lipschitz constant $K > 0$, that is,

$$\big|\big|f\big(z, \boldsymbol{X}(z)\big) - f\big(z, \boldsymbol{Y}(z)\big)\big|\big| \leq K ||\boldsymbol{X}(z) - \boldsymbol{Y}(z)||, \tag{2}$$

for all $\boldsymbol{X}, \boldsymbol{Y}$ in $\overline{U}$ and $a \leq z \leq b$. For each $i = 1, \dots, n$, let the coordinate functions $X_i(z)$ satisfy $m$ initial conditions at $z = a$ and the remaining $(n - m)$ conditions at $z = b$, that is,

$$(X_1, X_2, \dots, X_m)'(a) = (a_1, \dots, a_m)' \in \boldsymbol{R}^m, \tag{3}$$

$$\big(X_{\sigma(1)}, \dots, X_{\sigma(n-m)}\big)'(b) = (b_1, \dots, b_{n-m})' \in \boldsymbol{R}^{n-m} \tag{4}$$

where $\sigma: \{1, 2, \dots, n\} \to \{1, 2, \dots, n\}$ is a bijection.

The system $(1)-(4)$ defines a two-point BVP whose solution is not completely known at any of the two boundary points $z = a$ or $b$. Such BVPs are a common object of investigation in mathematical, physical, and engineering sciences. Most often, such BVPs do not possess a closed form solution although an approximate analytical solution may be found using topological methods such as the homotopy perturbation method. Consequently, one is forced to look for numerical solutions in order to unfold the inherent scientific information. In view of this, the differential equation in higher dimensions is converted to a nonlinear algebraic equation in higher dimensions by approximating $\boldsymbol{X}(z)$ with a Galerkin type expansion in terms of a suitable orthogonal basis. Such basis functions should satisfy the boundary conditions of the BVP in hand. The nonlinear algebraic equation is then solved numerically for the unknown coefficients with the help of an iterative scheme such as Newton-Raphson or bisection.

An easy numerical procedure to solve two-point BVPs such as the one described in (1)- (4) is the shooting method. Due to generality and applicability of the shooting technique in solving many different types of BVPs in ODEs, different shooting methods have been developed in the literature, based on the type of the two-point BVP in hand [1 - 9].

Although the shooting technique is a direct numerical approach towards solving nonlinear BVPs, several difficulties arise during its implementation. For example, a reasonably good guess for the unknown initial conditions is desired to be made at one of the boundaries. Otherwise, the underlying Newton iterates may not converge. Also, the location of the root within the radius of convergence is not known in advance. This complexity increases with $n$ and the use of shooting method may become impractical.

Ha [7] has discussed a shooting technique for nonlinear two-point BVPs and achieved rapid numerical convergence. Liu [10] has proposed Lie-group preserving schemes for integrating (1). Based on this, Liu [11] has developed a Lie-group shooting method for solving second order nonlinear BVPs numerically.

Using a modification of Adomian decomposition, Wazwaz [6] has developed an efficient shooting technique to handle higher order nonlinear two-point BVPs numerically.

In spite of plenty of available research on shooting techniques, their usage has been limited in the past. This serves as a motivation for the present chapter in order to highlight the power of the shooting techniques to efficiently solve higher order two-point BVPs numerically. More precisely, we mainly describe the nonlinear shooting method of Singh [17] for higher order two-point BVPs arising in physical situations.

The following well known results for systems of ODEs can be found in Birkhöff and Rota [12]. The reader interested in implementing the shooting method can skip these results for the time being as well as the content of Section 3, and may jump to study the methodology explained in Section 4. The theory part can be studied later if needed.

**Theorem 1**. *If $\boldsymbol{X} = \boldsymbol{A}(z)$, $\boldsymbol{B}(z)$ are any two solutions of* (1) *where $f\big(z, \boldsymbol{X}(z)\big)$ is continuous and satisfies Lipschitz condition of* (2), *then*

$$||\boldsymbol{A}(a+h) - \boldsymbol{B}(a+h)|| \leq e^{K|h|}||\boldsymbol{A}(a) - \boldsymbol{B}(a)||, 0 < h < b - a. \qquad (5)$$

**Corollary 2.** *Let $\boldsymbol{X}(z, \boldsymbol{c})$ be a solution of* (1) *with $\boldsymbol{X}(z, \boldsymbol{c}) = \boldsymbol{c}$ and the hypotheses of* Theorem 1 *are satisfied. Let $\boldsymbol{X}(z, \boldsymbol{c})$ exists on $|c - c_0| \leq k_1, |z - a| \leq k_2$. Then $\boldsymbol{X}(z, \boldsymbol{c})$ is a continuous function of z and c. Further, if $c \to c_0$, then $\boldsymbol{X}(z, \boldsymbol{c}) \to \boldsymbol{X}(z, \boldsymbol{c_0})$ uniformly on $|z - a| \leq k_2$.*

We recall that a smooth function $\boldsymbol{Y} \colon \boldsymbol{R} \to \boldsymbol{R}^n$ is said to be an approximate solution of the differential equation $d\boldsymbol{X}/d\boldsymbol{z} = f\big(z, X(z)\big), \boldsymbol{X}(0) = \boldsymbol{X}_0$, with an error at most $\eta$ if

$$||\boldsymbol{Y}(z) - \boldsymbol{X}(z)|| < \eta \text{ for all } z \in [a, b],$$

with deviation at most $\epsilon$ if $\boldsymbol{Y}(z)$ is continuous and satisfies the differential inequality

$$\big|\big|d\boldsymbol{Y}/dz - f\big(z, \boldsymbol{X}(z)\big)\big|\big| \leq \epsilon.$$

**Theorem 3.** (**Runge-Kutta Method**) *Let $\mathcal{P}: a = z_0 < z_1 < \cdots < z_{s-1} < z_s = b$ be a partition of the interval $[a, b]$. An approximate solution of the initial value problem*

$$\frac{dX(z)}{dz} = f(z, X(z)); X(a) = X_0,$$

*on $[a,b]$ is given by the following iterative scheme*

$$X^0 = X_0; X^{i+1} = X^i + \frac{h}{6}(k_1 + 2k_2 + 2k_1 + k_1); i = 0,1,2, ...,$$

*where $h = (b - a)/s$ and*

$$k_1 = f(z_i, X^i), k_2 = f(z_i + h/2, X^i + hk_1/2),$$

$$k_3 = f(z_i + h/2, X^i + hk_2/2), k_4 = f(z_i + h, X^i + hk_3),$$

*which is in an error $E = \mathcal{O}(h^5)$ at each iterative step.*

The following well known result is about the convergence of Newton iterates in higher dimensions (see for detail, Ortega [13], Tapia [14], Rall [15], and Gragg and Tapia [16]).

**Theorem 4.** *Let $A$ and $B$ be Banach spaces. Let $U$ be an open convex subset of $A$. Let $F: U \to B$ be such that map $F$ is Fréchét differentiable and satisfies*

$$\left| \left| \frac{dF}{dx} - \frac{dF}{dy} \right| \right| \le L||x - y||, for \ all \ x, y, \in U. \tag{6}$$

*For some $x_0 \in U$, assume that $F'(x_0)^{-1}$ is defined on all of $B$ and that $h := L\beta\eta \le 1/2$ where $||(dF/dx)(x_0)^{-1}|| \le \eta$ and $||(dF/dx)(x_0)^{-1}F(x_0)|| \le \eta$. Define*

$$t^* := \frac{1 - \sqrt{1 - 2h}}{\beta L}; t^{**} := \frac{1 + \sqrt{1 - 2h}}{\beta L}; S := \{x \in U \mid ||x - x_0|| \le t^*\}.$$

*Then sequence of Newton iterates $\{x_k\}$, where*

$$x_{k+1} = x_k - \frac{dF}{dx}(x_k)^{-1}F(x_k), k = 0,1,2 \ ... \tag{7}$$

*is well defined in $S$ and converge to a solution $x^*$ of $F(x^*) = 0$ which is unique in the set*

$$U \cap \{x \mid ||x - x_0|| < t^{**}\}.$$

*Moreover if $h < 1/2$, then the order of convergence is at least quadratic.*

The linear shooting method is described in Sec. 2. The technical detail about the nonlinear shooting technique is explained in Sec. 3. The numerical integration technique for the underlying BVP is then developed in Sec. 4 and is successfully implemented through examples in Sec. 5.

## 2. The linear shooting method

The linear shooting technique as we describe here converges rapidly for a wide class of linear BVPs in hand. The BVP (1)-(4) is called linear if the function $f(z, \boldsymbol{X}(z))$ in (1) is of the form $\boldsymbol{A}(z)\boldsymbol{X}(z) + \boldsymbol{B}(z)$, where $\boldsymbol{A}(z)$ is an $n \times n$ matrix function of $z$ and $\boldsymbol{B}(z) \in \boldsymbol{R}^n$ is a column matrix function of $z$. Each of $\boldsymbol{A}(z)$ and $\boldsymbol{B}(z)$ is independent of $\boldsymbol{X}(z)$. Consequently, the considered linear BVP can be described as follows:

$$\frac{d}{dz}\boldsymbol{X}(z) = \boldsymbol{A}(z) \cdot \boldsymbol{X}(z) + \boldsymbol{B}(z), z \in [a, b], \tag{8a}$$

$$(X_1, X_2, \dots, X_m)'(a) = (a_1, \dots, a_m)' \in \boldsymbol{R}^m, \tag{8b}$$

$$\left(X_{\sigma(1)}, \dots, X_{\sigma(n-m)}\right)'(b) = (b_1, \dots, b_{n-m})' \in \boldsymbol{R}^{n-m}. \tag{8c}$$

We follow the following steps of linear shooting technique. We construct the following set of linear IVPs

$$\frac{d}{dz}\boldsymbol{X}(z) = \boldsymbol{A}(z) \cdot \boldsymbol{X}(z) + \boldsymbol{B}(z), \boldsymbol{X}(a) = \xi_0 + e_{m+i}, \tag{9}$$

where $\xi_0 = (a_1, \dots, a_m, 0, 0, \dots, 0)' \in \boldsymbol{R}^n$, and $e_{m+i}, 1 \le i \le n - m$ is the element of the standard basis of $\boldsymbol{R}^n$ having its $(m+i)$-th entry 1 and each of the rest of the entries equal to zero. Using the theory of Wronskian, $(n - m)$ linearly independent solutions of (9) can be constructed, which remain linearly independent for all $z \in [a, b]$. Let $\boldsymbol{X}^{(1)}, \dots, \boldsymbol{X}^{(n-m)}$ be those $n - m$ linearly independent solutions of (9), such that

$$\boldsymbol{X}^{(i)}(z) = \left(X_1^{(i)}(z), \dots, X_n^{(i)}(z)\right)'; \boldsymbol{X}^{(i)}(a) = \xi_0 + e_{m+i}, 1 \le i \le n - m.$$

Then the solution of the BVP (8) can be expressed as a linear combination of the $n - m$ linearly independent column vectors $\boldsymbol{X}^{(1)}, \dots, \boldsymbol{X}^{(n-m)}$, that is, there exist scalar $\alpha_1, \dots, \alpha_{n-m}$ such that

$$\boldsymbol{X}(z) = \alpha_1 \boldsymbol{X}^{(1)}(z) + \cdots + \alpha_{n-m}\boldsymbol{X}^{n-m}(z). \tag{10}$$

In order that the vector $\boldsymbol{X}(z)$ in (10) satisfies the BVP (8), it must satisfy the remaining boundary condition at $z = b$. In view of this, we have

$$\alpha_1 X_{\sigma(i)}^{(i)}(b) + \cdots + \alpha_{n-m} X_{\sigma(i)}^{(i)}(b) = b_i, i = 1, \dots, n - m \tag{11}$$

which defines a system of $n - m$ linear algebraic equations in $\alpha_1, \dots, \alpha_{n-m}$ and hence can be solved using Gauss elimination method. Thus, using the Runge-Kutta method, one can compute a numerically solution of (8) which satisfies the condition (11) at the second boundary. It is easy to implement the aforementioned procedure in MATLAB using the subroutine `ode45` to solve the underlying $(n - m)$ IVPs at a discrete set of points in the interval $[a, b]$.

Linear BVPs arise frequently in hydrodynamic studies in fluid mechanics. In view of this, the linear shooting technique as explained here has been utilized in solving complex linear two-point BVPs in [18-21] arising from the Couette-Taylor instability and in [22] regarding the transient convection in ferrofluids.

### 3. The nonlinear shooting method

Following Singh [17], let $c \coloneqq (c_1, c_2, \ldots, c_{n-m})' \in R^{n-m}, n > m$, be independent of $z$ such that the smooth function $\boldsymbol{X}(z, c)$ satisfies (1) - (4), that is,

$$\frac{d}{dz}\boldsymbol{X}(z, c) = f\big(z, \boldsymbol{X}(z, c)\big); \ \boldsymbol{X}(a, c) = \boldsymbol{X}(a); \boldsymbol{X}(b, c) = \boldsymbol{X}(b). \tag{12}$$

Assume that the functions $X(z, c)$ and $f$ are defined on the closed rectangle $R$, where

$$R = [a, b] \times \overline{U} \subset [a, b] \times \boldsymbol{R}^{n-m},$$

and are continuously differentiable in $R \setminus \partial R$ such that $\partial f / \partial \boldsymbol{X}$ is nonsingular there. We observe that at point $(z, c)$, and each $i = 1, 2, \ldots n - m$, we have

$$\frac{d}{dz}\left(\frac{\partial \boldsymbol{X}}{\partial c_i}\right)(z, c) = \frac{\partial}{\partial c_i}\frac{d\boldsymbol{X}}{dz}(z, c) = \frac{\partial}{\partial c_i}f\big(z, \boldsymbol{X}(z, c_i)\big) = \frac{\partial f}{\partial \boldsymbol{X}}\big(z, \boldsymbol{X}(z, c_i)\big)\frac{\partial \boldsymbol{X}}{\partial c_i}(z, c). \tag{13}$$

If we let $\boldsymbol{X}(a, c) = (a_1, \ldots, a_m, c_1, c_2, \ldots, c_{n-m})'$, then we have

$$\frac{\partial \boldsymbol{X}}{\partial c_i}(a, c) = e_{m+i}.$$

Consequently, for each component $c_i$ of $c$, the derivative $\partial X / \partial c_i$ satisfies the following $(n - m)$ initial value problems (IVPs)

$$\frac{d}{dz}\left(\frac{\partial \boldsymbol{X}}{\partial c}\right)(z, c) = \frac{\partial f}{\partial \boldsymbol{X}}\big(z, \boldsymbol{X}(z, c)\big)\left(\frac{\partial \boldsymbol{X}}{\partial c}\right)(z, c); \left(\frac{\partial \boldsymbol{X}}{\partial c}\right)(a, c) = [e_{m+1}, \ldots, e_n]. \tag{14}$$

In view of the aforementioned discussion, we have the following result.

**Theorem 5 (Singh [17])**. *The hypothesis of Theorem* 4 *is satisfied by the sequence of functions* $\{F(c^k)\}$ *defined as follows:*

$$F(c^k) = \tilde{X}(b, c^k) - \tilde{X}(b), k = 0,1,2, \ldots \tag{15}$$

*on the closed rectangle* $\overline{U} \subset \boldsymbol{R}^{n-m}$ *containing the point* $c$ *in its interior, where*

$$\tilde{X}(z, c^k) = \left(X_{\sigma(1)}(z, c^k), \ldots, X_{\sigma(n-m)}(z, c^k)\right)' \text{ is in } [a, b] \times \overline{U}, \text{such that}$$

$$\tilde{X}(b) = \left(b_1, b_2, \ldots, b_{(n-m)}\right)'.$$

*The sequence* $\{c^k\}$ *is defined recursively by the following:*

$$c^0 = c, c^{k+1} = c^k - \left(\frac{\partial F}{\partial x}(c^k)\right)^{-1}F(c^k), k = 1,2, \ldots \tag{16}$$

For proof of Theorem 5, the reader is referred to consult Singh [17]. The following corollary is immediate from Theorem 5.

**Corollary 6 (Singh [17]).** *The sequence $\{c^k\}, k = 0,1, \ldots$ as in (16) is well defined and lies in the set $\{x \mid \|x - c\| < t^*\} \subset U_0$ and converges to unique solution $c^*$ of the equation $F(x) = 0$ on the set $U_0 \cap \{x \mid \|x - c\| < t^{**}\}$. Further, $\tilde{X}(z, c) \to \tilde{X}(z, c^*)$ as $c \to c^*$.*

**Remark.** The function defined by the sequence $\{F(c^k)\}$ as above, satisfies

$$\|F(c^{k+1}) - F(c^k)\| \le e^K \left| \left(dF/dc\,(c^k)\right)^{-1} F(c^k) \right|.$$

and hence, is dominated by the sequence $\{c^k\}$. To see this, consider the partition

$$a = z_0 < z_1 < \cdots < z_{N-1} < z_N = b, z_i = a + i(b - a)/N; 1 \le i \le N, \text{ of } [a, b].$$

Since $f$ satisfies Theorem 4 with Lipschitz constant $K$, we have the following or each fixed $k$ and $a = z_{i-1}$, and any two solutions $X(z, c^{k+1})$ and $X(z, c^k)$ of (12).

$$\|X(z_i, c^{k+1}) - X(z_i, c^k)\| \le e^{K(z_i - z_{i-1})} \|X(z_{i-1}, c^{k+1}) - X(z_{i-1}, c^k)\|$$

$$\cdots$$

$$\le e^{K(z_i - z_0)} \|X(z_0, c^{k+1}) - X(z_0, c^k)\|, \qquad (17)$$

and correspondingly, the following is satisfied.

$$\left|\|\tilde{X}(z_i, c^{k+1}) - \tilde{X}(z_i, c^k)\|\right| \le e^{K(z_i - z_{i-1})} \left|\|\tilde{X}(z_{i-1}, c^{k+1}) - \tilde{X}(z_{i-1}, c^k)\|\right|$$

$$\cdots$$

$$\le e^{K(z_i - z_0)} \left|\|\tilde{X}(z_0, c^{k+1}) - \tilde{X}(z_0, c^k)\|\right|. \qquad (18)$$

From (18) we obtain the desired assertion for $F(c^k), k = 0,1 \ldots$ as follows:

$$\|F(c^{k+1}) - F(c^k)\| = \|\tilde{X}(1, c^{k+1}) - \tilde{X}(1, c^k)\|$$

$$\le e^K \|\tilde{X}(0, c^{k+1}) - \tilde{X}(0, c^k)\|$$

$$= e^K \|c^{k+1} - c^k\| = e^K \left| \left(\partial F/\partial c\,(c^k)\right)^{-1} F(c^k) \right|,$$

since the sequence $\{\partial F/\partial c(c^k)\}$ is constant. Also, $F(c^k) \to F(c^*) \to 0$ for $c^k \to c^*$.

## 4. The Numerical scheme for nonlinear shooting method

In this section, we develop a numerical scheme for applying the nonlinear shooting method described in Section 3. In view of Theorem 5, an arbitrary choice of the Lipschitz constant $\kappa > 0$ ensures at least quadratic convergence of the underlying Newton iterates. The underlying iterative scheme involves computation of an approximate numerical solution $X(z, c^*)$ of (1)-(4) which at $(k + 1)$-th iteration involves simultaneous numerical integration of (12) from $z = a$ to $z = b$ to compute $F(c^k)$ followed by a numerical integration of IVPs defined by (14) for each $c_i$ to numerically compute $\frac{\partial X}{\partial c}(b, c^k)$. The associated numerical integrations are to be performed on the interval $[a, b]$. In this view, we consider the following system of IVPs:

$$\frac{d}{dz}\begin{pmatrix} \boldsymbol{X} \\ \frac{\partial \boldsymbol{X}}{\partial c_i^k} \end{pmatrix}(z, c^k) = \begin{pmatrix} f\left(z, \boldsymbol{X}(z, c^k)\right) \\ \frac{\partial f}{\partial \boldsymbol{X}}\frac{\partial \boldsymbol{X}}{\partial c_i^k}(z, c^k) \end{pmatrix}; \begin{pmatrix} \boldsymbol{X} \\ \frac{\partial \boldsymbol{X}}{\partial c_i^k} \end{pmatrix}(a, c^k) = \begin{pmatrix} \xi^k \\ e_{m+i} \end{pmatrix}, \tag{19}$$

for each $i = 1, \ldots, n - m$, where $\xi^k = \left(a_1, \ldots, a_m, c_1^k, c_2^k, \ldots, c_{n-m}^k\right)'$. So, from (19), we have $(n - m)$ IVPs in ODEs, each of which is of order $2n$ and can be integrated numerically. We explain the procedure in the following steps.

**Step 1.** Using the fourth order Runge-Kutta method or the more efficient Runge-Kutta Fehlberg method, we solve the $(n - m)$ IVPs as in (19) for a finite set of points in the interval $[a, b]$, to obtain the solution at $z = b$, that is,

$$\begin{pmatrix} \boldsymbol{X} \\ \frac{\partial \boldsymbol{X}}{\partial c_i^k} \end{pmatrix}(b, c^k), i = 1, \ldots, n - m; \ c^k = \left(c_1^k, \ldots, c_{n-m}^k\right), k = 0, 1, 2, \ldots,$$

with an appropriate step size $h$ starting with an initial guess $c^0$ for $k = 0$.

**Step 2.** In the next step we extract the vector $\tilde{X}(b, c^k) = \left(X_{\sigma(1)}(b, c^k), \ldots, X_{\sigma(n-m)}(b, c^k)\right)'$ and the Jacobian matrix $\frac{\partial \tilde{X}}{\partial c}(b, c^k)$ from the numerically computed solution in the preceding step so that we have

$$\frac{\partial \tilde{X}}{\partial c}(b, c^k) = \left(\frac{\partial \tilde{X}}{\partial c_1^k}(b, c^k), \frac{\partial \tilde{X}}{\partial c_2^k}(b, c^k), \cdots, \frac{\partial \tilde{X}}{\partial c_{n-m}^k}(b, c^k)\right). \tag{20}$$

Suppose the numerical solution is sorted within the specified tolerance tol., then this amounts to check if the following condition is satisfied

$$\left\| \tilde{X}(b, c^k) - \tilde{X}(b) \right\| < \text{tol.} \tag{21}$$

so that at the $k$-th iteration, $\tilde{X}(z, c^k) \approx X(z)$ is the desired approximated solution within an error not exceeding tol. Otherwise, we go to the next step for implementing the next Newton's iteration.

**Step 3.** If $\left\| \tilde{X}(b, c^k) - \tilde{X}(b) \right\| \geq$ tol., then the $(k+1)$-th Newton iteration is performed to obtain $c^{k+1}$ by solving the following system of $(n - m)$ linear algebraic equations in $c_1^{k+1}, \ldots, c_{n-m}^{k+1}$ given by

$$\frac{\partial \tilde{X}}{\partial c}(b, c^k)c^{k+1} = \frac{\partial \tilde{X}}{\partial c}(b, c^k)c^k - \{\tilde{X}(b, c^k) - \tilde{X}(b)\}. \tag{22}$$

We use Gauss-elimination method in order to solve the linear system in (22) which is easy to implement in computer programs and is free from any type of numerical singularity.

With the computation of $c^{k+1}$ using (22) one iteration of the shooting method is completed. The whole procedure is repeated through **Step 1-3** to obtain $c^{k+2}$ till the numerically computed solution is within the specified tolerance, that is, it satisfies (21).

## 5. Numerical examples and discussion

It is easy to write a program in MATLAB for simultaneously solving (20) and (22) recursively, till the scheme converges and meets the desired tolerance. The numerical codes for implementing the present numerical scheme can be easily written in MATLAB by following the excellent text of Mathews and Fink [23]. To justify applicability of the proposed numerical procedure of nonlinear shooting technique, we consider the following examples.

**Example 1.** Consider the following second-order nonlinear BVP

$$\frac{d}{dz}\begin{pmatrix} X_1 \\ X_2 \end{pmatrix} = \begin{pmatrix} X_2 \\ 2X_1^3 - 6X_1 - 2z^3 \end{pmatrix}; X_1(1) = 2, X_1(2) = 5/2,$$

for $z \in [1,2]$. Here $n = 2, m = 1, c = c_1, a = 1, b = 2,$ $\sigma$ is the identity permutation of $\{1,2\}$ so that $\bar{X}(z,c) = X_1(z,c)$, and

$$f(z, \boldsymbol{X}(z)) = \begin{pmatrix} f_1 \\ f_2 \end{pmatrix}(z, \boldsymbol{X}(z)) = \begin{pmatrix} X_2 \\ 2X_1^3 - 6X_1 - 2z^3 \end{pmatrix}.$$

We construct the approximate solution $\boldsymbol{X}(z,c) = \big(X_1(z,c), X_2(z,c)\big)'$ so that $\boldsymbol{X}(z,c)$ and $\frac{\partial \boldsymbol{X}}{\partial c}(z,c)$ satisfying the following

$$\frac{d\boldsymbol{X}}{dz}(z,c) = f\big(z, \boldsymbol{X}(z,c)\big); \boldsymbol{X}(1,c) = \begin{pmatrix} 2 \\ c_1 \end{pmatrix}$$

$$\frac{d}{dz}\frac{\partial \boldsymbol{X}}{\partial c_1}(z,c) = \begin{pmatrix} \dfrac{\partial f_1}{\partial X_1} & \dfrac{\partial f_1}{\partial X_2} \\ \dfrac{\partial f_2}{\partial X_1} & \dfrac{\partial f_1}{\partial X_1} \end{pmatrix} \frac{\partial \boldsymbol{X}}{\partial c_1}(z,c) = \begin{pmatrix} 0 & 1 \\ 6X_1^2 - 6 & 0 \end{pmatrix} \frac{\partial \boldsymbol{X}}{\partial c_1}(z,c),$$

$$\frac{\partial \boldsymbol{X}}{\partial c_1}(1,c) = \begin{pmatrix} 0 \\ 1 \end{pmatrix} = e_2.$$

In view of (19), we need to solve the following IVP numerically using Runge-Kutta method.

$$\frac{d}{dz}\begin{pmatrix} X_1 \\ X_2 \\ \frac{\partial X_1}{\partial c_1} \\ \frac{\partial X_2}{\partial c_1} \end{pmatrix} = \begin{pmatrix} X_2 \\ 2X_1^3 - 6X_1 - 2z^3 \\ \frac{\partial X_2}{\partial c_1} \\ 6(X_1^2 - 1)\frac{\partial X_1}{\partial c_1} \end{pmatrix}; \begin{pmatrix} X_1 \\ X_2 \\ \frac{\partial X_1}{\partial c_1} \\ \frac{\partial X_2}{\partial c_1} \end{pmatrix}(1, c) = \begin{pmatrix} 2 \\ c_1 \\ 0 \\ 1 \end{pmatrix}, \qquad (23)$$

where $c_1$ is the

Now applying the **Steps 1-3** as explained in the preceding section, we compuet the numerical solution of the given BVP. The exact solution is $X_1(z) = z + \frac{1}{z}$, $1 \leq z \leq 2$.

We choose a step size of $h = 0.02$ to numerically integrate the IVP (23) using Runge-Kutta method within a tolerance of $10^{-6}$. To apply Runge-Kutta method, an initial guess for $c_1^0$ is needed, which can be specified as shown Table 1, where the convergence analysis of the solution is also shown for different initial guesses for $c_1^0$. Clearly, $c_1^0 = 0$ is appropriate corresponding to which the numerical convergence is achieved for first iteration within an error of the order of $10^{-8}$. Larger the value of $|c_1^0|$, larger is the number of iterations required to have the numerical convergence within an error at most of the order of $10^{-8}$.

**Table 1.** Variation of $k$ with $c_1^0$ for a step size of 0.02.

| $c_1^0$ | $k$ | $c_1^k$ | $\|\|\tilde{X}(2, c^k) - \tilde{X}(2)\|\|$ |
|---|---|---|---|
| -100 | 118 | $-5.567960e-09$ | $2.051692e-13$ |
| -10 | 49 | $-5.547987e-09$ | $2.462031e-10$ |
| -1.0 | 34 | $-5.140315e-09$ | $5.266837e-09$ |
| 0.0 | 01 | $0.000000e-00$ | $6.857203e-08$ |
| 0.1 | 05 | $-5.554485e-09$ | $1.661671e-10$ |
| 0.5 | 13 | $-5.567931e-09$ | $5.559997e-13$ |
| 1.0 | 32 | $-4.726264e-09$ | $1.036603e-08$ |
| 5.0 | 58 | $-5.567060e-09$ | $1.131273e-11$ |
| 10.0 | 62 | $-4.687076e-09$ | $1.084867e-08$ |
| 20.0 | 79 | $-5.544791e-09$ | $2.855325e-10$ |
| 50.0 | 98 | $-5.276342e-09$ | $3.591604e-09$ |

The Example 1 was considered by Ha [7] in which no numerical solution was found to exist with his numerical method for $c_1^0 \geq 1$. On the other hand, it is clear from Table 1 that the present numerical scheme converges even for $c_1^0 \geq 1$. Moreover, for $c_1^0 = 0.6975$, the present numerical scheme converges in 20 iterations to the approximate numerical solution within an error of at most $3.483020e - 07$, whereas Ha's numerical method convergence only after 9982 iterations.

The system of ODEs in the Example 1 is locally Lipschitz with the Lipschitz constant $\kappa \geq 40$ in the following set

$$R = \{(X_1, X_2) \in \mathbf{R}^2 : |X_1 - 2| \leq 0.5\}.$$

By Theorem 5, the present numerical scheme converges for every value of $c_1^0$ taken from the set $R$ on the line $X_1 = 2$. However, for larger values of $c_1^0$, the rate of convergence may slow down as can be seen from Table 1. The numerically computed profiles of $X_1$ and $X_2$ with $z$ are shown in Fig. 1, where the solid lines denote the actual solution and the asterisks ✳ denote the numerically computed values. It is evident from the Fig.1 that the graph of numerically approximated solution using the present method is indistinguishable from that of the closed form solution.

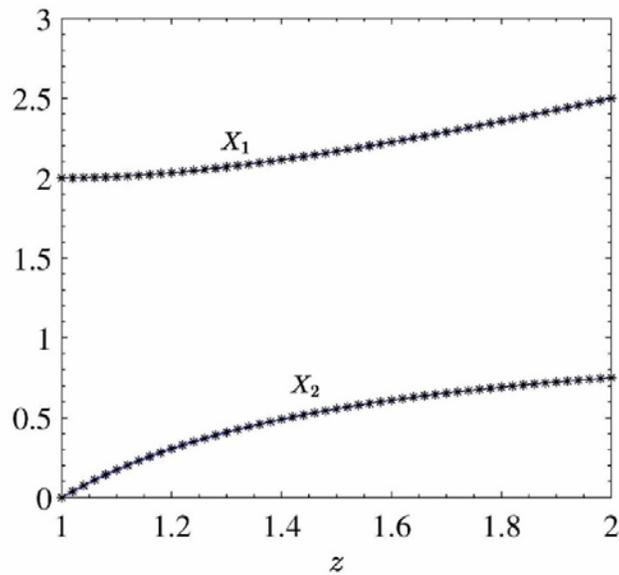

**Figure 1**. Numerically computed solutions (✳) vs the closed form solution (-) of the BVP considered in Example 1(See Singh [17]).

In fact, the present numerical method will converge for all initial guesses in the BVPs originating from locally Lipschitz ODEs, where the Jacobian is computed only once for all iterations, thereby lowering the expenses of the numerical computations.

**Example 2.** Consider the following fifth order coupled nonlinear BVP arising in the boundary layer flow due to a stretching sheet under temperature variations.

$$f''' + ff'' - (f')^2 = 0; f(0) = 0, f'(0) = 1 = \theta(0), \qquad (24a)$$

$$\theta'' + \text{Pr}\,\theta'\, f = 0; f'(5) = 0 = \theta(5). \tag{24b}$$

If we let $f = X_1$, $f' = X_2$, $f'' = X_3$, $\theta = X_4$, and $\theta' = X_5$, then $\boldsymbol{X}(z) = (X_1, X_2, X_3, X_4, X_5)'$

and the given nonlinear system (24) reduces to the following equivalent system

$$\frac{d}{dz}\begin{pmatrix} X_1 \\ X_2 \\ X_3 \\ X_4 \\ X_5 \end{pmatrix} = f\big(z, \boldsymbol{X}(z)\big) = \begin{pmatrix} X_2 \\ X_3 \\ -X_1 X_3 + (X_2)^2 \\ X_5 \\ -\text{Pr}\,X_1 X_5 \end{pmatrix}, 0 \le z \le 5, \tag{25a}$$

along with the following boundary conditions given by

$$X_1(0) = 0, X_2(0) = 1 = X_4(0); X_2(5) = 0 = X_4(5). \tag{25b}$$

Here, $n = 5, m = 2, c = (c_1, c_2)', a = 0, b = 5$, and $\sigma$ is the bijection with

$$\sigma(1) = 2, \sigma(2) = 4, \tag{25b}$$

so that $\tilde{X}(z, c) = (X_2(z, c), X_4(z, c))'$. As before, let the approximate solution of (25) be

$$\boldsymbol{X}(z, c) = (X_1(z, c), X_2(z, c), X_3(z, c), X_4(z, c), X_5(z, c))'$$

so that $\boldsymbol{X}(z, c)$ and $\frac{\partial \boldsymbol{X}}{\partial c}(z, c)$ satisfy the following IVP:

$$\frac{d\boldsymbol{X}}{dz}(z, c) = f\big(z, \boldsymbol{X}(z, c)\big), \tag{26a}$$

$$\frac{d}{dz}\frac{\partial \boldsymbol{X}}{\partial c_i}(z, c) = \begin{pmatrix} 0 & 1 & 0 & 0 & 0 \\ 0 & 0 & 1 & 0 & 0 \\ -X_3 & 2X_2 & -X_1 & 0 & 0 \\ 0 & 0 & 0 & 0 & 1 \\ -\text{Pr}\,X_5 & 0 & 0 & 0 & -\text{Pr}\,X_1 \end{pmatrix}\frac{\partial \boldsymbol{X}}{\partial c_i}(z, c), \tag{26b}$$

$$\boldsymbol{X}(0, c) = \begin{pmatrix} 0 \\ 1 \\ c_1 \\ 1 \\ c_2 \end{pmatrix}; \frac{\partial \boldsymbol{X}}{\partial c}(0, c) = \begin{pmatrix} 0 & 0 \\ 0 & 0 \\ 1 & 0 \\ 0 & 0 \\ 0 & 1 \end{pmatrix} = (e_3, e_5), \tag{26c}$$

where $c_1$ and $c_2$ are the unknown missing boundary conditions to be approximated numerically using the shooting technique. So, we follow the **Steps 1-3** of Sec. 4, to solve the IVPs defined by (26) numerically for a step size of 0.01 and a specified tolerance of $10^{-6}$. The missing boundary conditions for each case $\text{Pr} = 0.71$, 1, 6 as obtained using the present numerical procedure are given in Table 2, where we find that for the nonlinear BVP (24), $c_1$ is independent of the parameter Pr while $c_2$ changes on changing Pr.

**Table 2.** Variation of the number of iterations $k$ with Pr for the convergence within the specified tolerance.

| Pr | $k$ | $c_1; c_2$ | $\|\|\bar{X}(5, c^k) - \bar{X}(5)\|\|$ |
|------|-----|------------------------------|---------------------------------|
| 0.71 | 19 | $-1.001396e + 00; -4.755625e - 01$ | $8.784541e - 07$ |
| 1.0  | 16 | $-1.001396e + 00; -5.872225e - 01$ | $7.851039e - 07$ |
| 6.0  | 8  | $-1.001396e + 00 ; -1.738095e + 00$ | $2.235719e - 07$ |

The numerical scheme converges rapidly to the solution of the given BVP (24) within an error of the order of $10^{-7}$. Figure 2 shows the numerically computed solution for the considered numerical values of the parameter Pr. The profiles for $X_1$ and $X_2$ are independent of Pr, while the profile for $X_4$ vary significantly on varying Pr. This can be understood physically since $X_4 = \theta$, the temperature of the boundary layer in the vicinity of the stretching sheet surface, which has to vary with changing Pr, the Prandtl number of the viscous fluid under consideration.

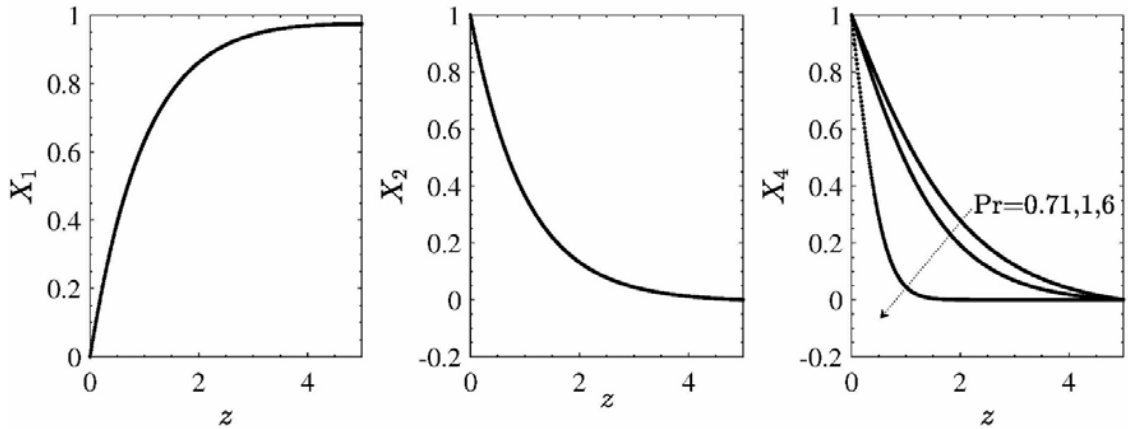

**Figure 2.** Plots of the numerical solution with $z$ using the nonlinear shooting method for various values of Pr.

Many other similar nonlinear high order BVPs can be handled efficiently with ease using the present numerical technique. Below, we provide some sample problems for the purpose; the first three problems have been taken from Vleggaar [24].

- $2f'''(\eta) + f(\eta)f''(\eta) = 0; \quad f(0) = 0, f'(0) = 1 = \theta(0),$

  $2\theta''(\eta) + \Pr \theta'(\eta)f(\eta) = 0; f'(5) = 0 = \theta(5).$

- $\eta f''' + (f - 1)(f'' - f'/\eta) = 0; f(1) = 1/2, f'(1) = 1 = \theta(1),$

  $\theta'' + \theta'(1 + \Pr f) = 0; \ f'(5) = 0 = \theta(5).$

- $\eta f''' + (f-1)(f'' - f'/\eta) - (f')^2 = 0; f(1) = 1/2, f'(1) = 1 = \theta(1),$

$$\theta'' + \theta'(1 + \Pr f)/\eta = 0; f'(5) = 0 = \theta(5).$$

- Solve $y'' = 2yy', 0 \leq t \leq 1; y(0) = 0, y(1) = 2$ by taking initial guess $\alpha_0 = 0, 0.5, 1$. What do you observe?

- Solve the following BVP numerically:

$$y'' = 400y + 400\cos^2(\pi x) + 2\pi^2\cos(2\pi x), y(0) = 0 = y(1).$$

## 6. Concluding remarks

The linear shooting method is discussed in brief to solve linear two-point BVPs in ODEs. A good number of references is provided to encourage the reader about its usage.

The nonlinear shooting method for solving nonlinear higher order BVPs is discussed in detail and the various steps involved in its implementation are described through concrete examples. Based on the analysis given in Singh [17], we have described a numerical scheme for obtaining numerical solutions of higher order nonlinear two-point BVPs arising in different physical domains within the framework of ODEs. The numerical procedure is easy to implement in standard computer packages, such as MATLAB. The underlying numerical scheme is found to converge rapidly to a solution of the nonlinear BVP in hand. In view of the ease of applicability and high rate of convergence of the proposed linear and nonlinear shooting techniques, we recommend their use for handling a wide variety of linear and nonlinear BVPs, originating in diverse areas of science and technology.

## References

[1]  H. B. Keller. "Numerical methods for two-point boundary value problems." *Soc. Indust. Appl. Math.,* Vol. **24**, pp. 1−61, 1976.

[2]  A. Granas, R. B. Guenther, and J. W. Lee. "The shooting method for the numerical solution of a class of nonlinear boundary value problems." *SIAM J. Numer. Anal.,* Vol. **16**:5, pp. 828−836, 1979.

[3]  R. M. M. Mattheij and G. W. M. Staarink. "On optimal shooting intervals." *Math. of Comp.,* Vol. **42**:165, pp. 25−40, 1984.

[4]  J. Stoer and R. Bulirsch. *Introduction to Numerical Analysis*. Springer-Verlag, New York, 1993.

[5]  M. E. Kramer and R. M. M. Mattheij. "Application of global methods in parallel shooting." *SIAM J. Numer. Anal.,* Vol. **30**:6, pp. 1723−1739, 1993.

[6]  A. M. Wazwaz. "Approximate solutions to boundary value problems of higher order by modified decomposition method." *Comp. and Math. with Appl.,* Vol. **40**, pp. 679−69, 2000.

[7]  Sung. N. Ha. "A nonlinear shooting method for two-point boundary value problems." *Comp. Math. Appl.,* Vol. **42**, pp. 1411–1420, 2001.

[8]     A. M. Wazwaz. "A reliable algorithm for obtaining positive solutions for nonlinear boundary value problems." *Comp. and Math. with Appl.,* Vol. **41**, pp. 1237–1244, 2001.

[9]     B. S. Attili and M. I. Syam. "Efficient shooting method for solving two-point boundary value problems." *Chaos, Soliton and Fractals.,* Vol. **35**, pp. 895–903, 2008.

[10]    C. S. Liu. "Cone of non-linear dynamical system and group preserving schemes." *Int. Jour. Nonlinear Mech.,* Vol. **36**, pp. 1047–1068, 2001.

[11]    C. S. Liu. "The Lie-group shooting method for boundary-layer problems with suction/injection/reverse flow conditions for power-law fluids." *Int. Jour. Nonlinear Mech.,* Vol. **46**, pp. 1001–1008, 2011.

[12]    G. Birkhoff and G. C. Rota. *Ordinary differential equations.* John Wiley and Sons, New York, 1978.

[13]    J. M. Ortega. "The Newton-Kantorovich theorem." *The Amer. Math. Monthly.,* Vol. **75**:6, pp. 658–660, 1968.

[14]    R. A. Tapia. "The Kantorovich theorem for Newton's method." *The Amer. Math. Monthly.,* Vol. **78**:6, pp. 389–392, 1971.

[15]    L. B. Rall. "A note on the convergence of Newton's method." *SIAM J. Numer. Anal.,* Vol. **11**:1, pp. 34–36, 1974.

[16]    W. B. Gragg and R. A. Tapia. "Optimal error bounds for the Newton-Kantorovich theorem." *SIAM J. Numer. Anal.,* Vol. **11**:1, pp. 10–13, 1974.

[17]    J. Singh. "A nonlinear shooting method and its application to nonlinear Rayleigh-Bénard convection." *ISRN Mathematical Physics.,* Vol. **2013,** Article ID 650208, 2013.

[18]    J. Singh and R. Bajaj. "Couette flow in ferrofluids with magnetic field." *J. Magnetism and Magnetic Materials.,* Vol. **294**, pp. 53–62, 2005.

[19]    J. Singh and R. Bajaj. "Stability of non-axisymmetric ferrofluid flow in rotating cylinders with magnetic field." *International Journal of Mathematics and Mathematical Sciences.,* Vol. **23**, pp. 3727–3737, 2005.

[20]    J. Singh and R. Bajaj. "Stability of ferrofluid flow in rotating porous cylinders with radial flow." *Magnetohydrodynamics.,* Vol. **42**, pp. 46–56, 2006.

[21]    J. Singh and R. Bajaj. "Non-axisymmetric modes of Couette-Taylor instability in ferrofluids with radial flow." *Magnetohydrodynamics.,* Vol. **42**, pp. 57–68, 2006.

[22]    J. Singh. "Energy relaxation for transient convection in ferrofluids." *Physical Review E.,* Vol. **82**, pp. 026311, 2010.

[23]    J. H. Mathews and K. D. Fink, *Numerical Methods using MATLAB*, Pearson Prentice Hall, New Jersey USA, 2004.

[24]    J. Vleggaar. "Laminar boundary-layer behaviour on continuous, accelerating surfaces." *Chemical Engg. Sci.,* Vol. **32**, pp. 1517−1525, 1977.